\documentclass[journal]{IEEEtran}

\usepackage{graphicx}
\usepackage{epsfig}
\usepackage{dcolumn}
\usepackage{bm}
\usepackage{amsmath}
\usepackage{amssymb}

\begin{document}
\title{Neuromorphic, Digital and Quantum Computation with Memory Circuit Elements}

\author{Yuriy~V.~Pershin
        and Massimiliano~Di~Ventra 
\thanks{Yu. V. Pershin is with the Department of Physics
and Astronomy and USC Nanocenter, University of South Carolina,
Columbia, SC, 29208 \newline e-mail: pershin@physics.sc.edu.}
\thanks{M. Di Ventra is with the Department
of Physics, University of California, San Diego, La Jolla,
California 92093-0319 \newline e-mail: diventra@physics.ucsd.edu.}
}

\maketitle

\begin{abstract}
Memory effects are ubiquitous in nature and the class of memory
circuit elements - which includes memristive, memcapacitive and
meminductive systems - shows great potential to understand and simulate the
associated physical processes. Here, we show that such elements can also be used in
electronic schemes mimicking biologically-inspired computer architectures, performing digital
logic and arithmetic operations, and can expand the capabilities of certain quantum computation schemes. In particular, we will discuss some
examples where the concept of memory elements is relevant to the realization
of associative memory in neuronal circuits, spike-timing-dependent plasticity of synapses, digital and
field-programmable quantum computing.
\end{abstract}


\section{Introduction}

\IEEEPARstart{M}{emristive}~\cite{chua71a,chua76a}, memcapacitive and meminductive~\cite{diventra09a} systems
constitute an important class of two-terminal circuit elements whose basic characteristics - namely, their resistance, capacitance and inductance - retain memory of the past states through
which the systems have evolved. The memory
features of these systems
are related to corresponding internal states of these elements
(e.g., atomic structure \cite{Yang08a}, spin polarization \cite{pershin08a,wang09a}, etc.) which can be influenced by an external control
parameter like the voltage, charge, current or flux.

Mathematically, $n$th-order $u$-controlled memory elements  are
defined by the relations~\cite{diventra09a}
\begin{eqnarray}
y(t)&=&g\left(x,u,t \right)u(t) \label{Geq1}\\ \dot{x}&=&f\left(
x,u,t\right) \label{Geq2}
\end{eqnarray}
where $x$ denotes a set of $n$ state variables describing the
internal state of the system, $u(t)$ and $y(t)$ are any two
fundamental circuit variables (i.e.,
current, charge, voltage, or flux) denoting input and output of
the system, $g$ is a generalized response, and $f$ is a continuous
$n$-dimensional vector function.  The different memory elements are determined
by three pairs of circuit variables: current-voltage (memristive systems),
charge-voltage (memcapacitive systems), and flux-current (meminductive systems).
Two other pairs (charge-current and voltage-flux) are linked
through equations of electrodynamics, and therefore do not give rise to any new
element. Devices defined by the relation of charge and
flux (which is the time integral of the voltage) are not considered as
a separate group since such devices can be redefined in the
current-voltage basis~\cite{chua71a}. Moreover, we note that
memristors, memcapacitors and meminductors are ideal (and rare) instances of memristive,
memcapacitive and meminductive systems, respectively (see Refs. \cite{chua71a,diventra09a,pershin11d}).
In this work, the terms memristors, memcapacitors and meminductors are reserved only for such ideal cases.

These memory systems turn out to be of very general interest in science and engineering and are potentially useful not just in information storage but also in apparently different areas of research
\cite{Kuekes05a,driscoll09b,pershin09b,Lehtonen09a,driscoll09a,Borghetti09a,Gergel09a,Lai10a,pershin10c,pershin10d,borghetti10a,jo10a,Driscoll10a,pershin11d}.
Of equal importance and arguably the less studied of all these applications so far, memory elements may be of use in the three
different paradigms of computation: analog, digital, and quantum. This is the subject of this paper.

In the case of analog computation, it was recently shown that
electronic circuits with memory circuit elements \cite{diventra09a} can
simulate processes typical of biological systems such as the adaptive
behavior of unicellular organisms \cite{pershin09b}, learning and associative memory~\cite{pershin10c}.
The concept of ``learning circuits'' has also been implemented recently using VO$_2$ as memory element~\cite{Driscoll10a}. Here, we first briefly discuss
existing memristive models of neural computing as well as realization of spike-timing-dependent plasticity (STDP) with first-order memristive systems.
Then, we formulate a model of a second-order memristive synapse which resembles closer the operation of its biological counterpart.

The implementation of classical logic operations with memristive systems was discussed theoretically in the past (see e.g. Refs. \cite{strukov05a,Snider07a,Lehtonen09a}) and demonstrated experimentally \cite{borghetti10a}. In this paper, we suggest a different scheme for classical logic operations utilizing a combination of memcapacitive and memristive systems.
Our scheme realizes all basic logic operations (NOT, OR, AND) as well as many other
operations (XOR, multi-input OR, AND, etc.) in a much simpler
way as compared to the alternative approaches~\cite{Lehtonen09a,borghetti10a}.
In addition, it provides a significant speed-up of operations. In this paper, we demonstrate experimentally for the first time
addition of two one-bit numbers using memory circuit elements. Our scheme requires only 16 steps compared to 87 steps as suggested in a previous approach \cite{Lehtonen09a}. Also, it is worth
stressing that in these cases computing {\it and} memory are integrated in the {\it same} platform, which is a major conceptual departure from present-day computing technology where memory and computing are physically
disjoint.

Finally, in the case of quantum computation we discuss specific quantum computing schemes where meminductive and memcapacitive systems could be used to generate an essentially infinite
number of programmable interaction Hamiltonians between two or more qubits inductively or capacitively coupled, in a manner we could term {\it field-programmable quantum computation}, the quantum analog of the well-established
classical field-programmable gate array. This opens up the possibility of expanding both our fundamental understanding of circuit quantum electrodynamics~\cite{Blais04a,Wallraff04a} as well as the optimization of quantum computing algorithms.

\section{Neural computation with memristive systems}

Neuromorphic computing circuits are designed by borrowing principles
of operation typical of the human (or animal) brain and, therefore, due to their intrinsic analog capabilities they
can potentially solve problems that are cumbersome (or outright intractable) by digital  computation. Examples of such problems include - but are not limited to - adaptive behavior, learning by association,  pattern recognition, fuzzy logic, etc.
Certain realizations of memristive systems can be very useful in such
circuits because of their intrinsic properties which
mimic to some extent the behavior of biological synapses. In addition, the most recent experimental
realizations of memory elements pertain to systems of dimensions in the nanometer regime \cite{Yang09b,Jo09a}, thus allowing a possible scale-up
of such elements in a chip to the number density of a typical human brain
(consisting of about 10$^{11}$ synapses/cm$^3$). In the following we first briefly review our work on {\it memristive neural networks} (MNNs) showing a practical realization of the
Hebbian learning, which states, in a very simplified form, that ``neurons that
fire together, wire together".

We then discuss one of the most important functions of synapses namely their spike-timing-dependent plasticity
\cite{Levy83a,Markram97a,Bi98a,Froemke02a}. The existing experimental approaches to STDP are based on relatively large-size VLSI circuits \cite{plast1,plast2,plast3,plast4},
three terminal transistor-like structures based on ionic conduction \cite{Lai10a} or combination of
memristive systems with CMOS elements \cite{jo10a}. Below, we discuss two different memristive system-based schemes in which
STDP can be realized. In the first, we may call ``bipolar'', first-order memristive systems probed by two composite pulses overlapping in time develop STDP. This is similar to what
has been studied in \cite{snider08a,Afifi09a,Barranco09b,parkin10a}. In the second realization, we could call ``intrinsic'', we provide a model of a memristive system which at the prize of increasing the number of state
variables to two, does show STDP even if the excitation pulses do not overlap temporally. This situation is closer to the actual operation of a biological synapse in which each synaptic event triggers a cascade of internal decaying processes/reactions.

\subsection{Memristive neural networks and Hebbian learning}

\begin{figure}[h]
 \begin{center}
\includegraphics[angle=0,width=6.5cm]{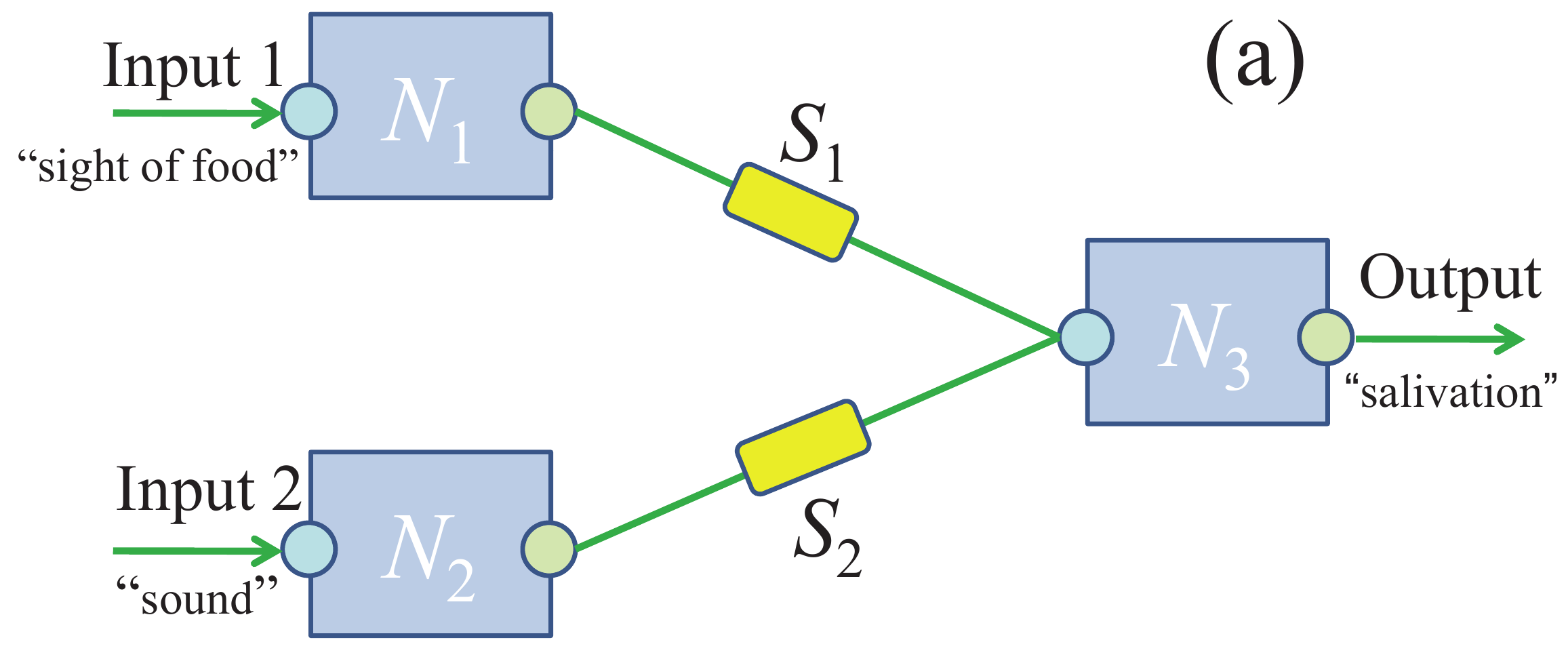}
\includegraphics[angle=0,width=6.5cm]{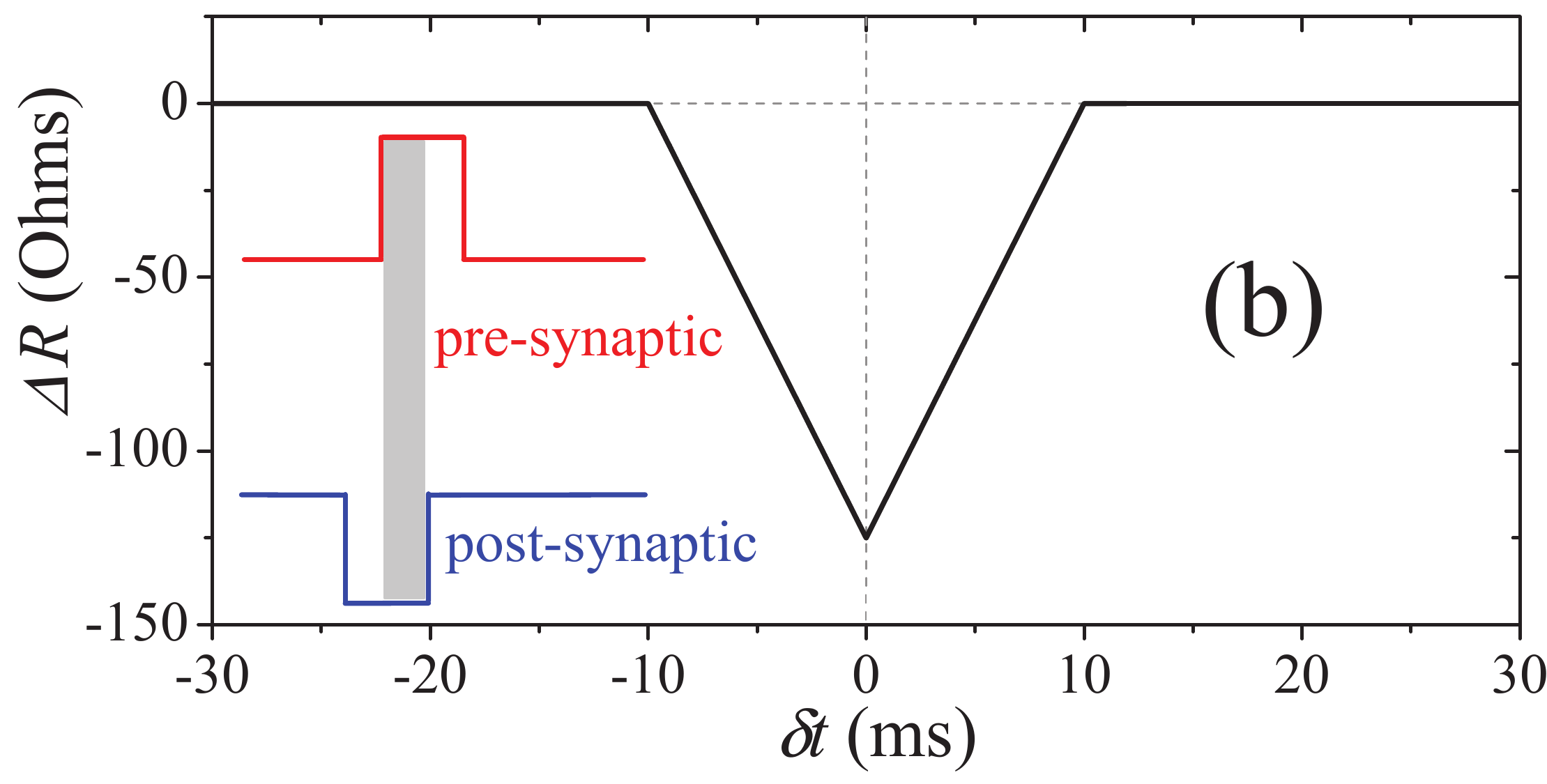}
\caption{(Color online) (a) A simple neural network containing three electronic neurons (N$_1$, N$_2$ and N$_3$) connected by two memristive synapses (S$_1$ and S$_2$). Reprinted from Ref. \cite{pershin10c}, copyright (2010), with permission from Elsevier. (b) Change in memristance $\Delta R$ induced by a pair of square pulses of opposite polarity applied to terminals of a first-order memristive system described by Eqs. (\ref{eq3}, \ref{eq4}) plotted as a function of a time interval between the forward (pre-synaptic) and backward (post-synaptic) pulses $\delta t =t_{post}-t_{pre}$. The pulses' form is shown schematically
 in the inset (not to scale). The grey rectangle designates the time window in which the memristance change occurs.
 The maximum pulse overlap is at $\delta t=0$ and corresponds to the maximum of $\Delta R$. This plot was obtained as a numerical
 solution of Eqs. (\ref{eq3}, \ref{eq4}) with $\alpha=0$, $\beta=-25$kOhms/(V$\cdot$s), $V_t=1.5$V, $R(t=0)=2$kOhms, $R_1=1$kOhms and $R_2=10$kOhms. Pre-synaptic and post-synaptic pulses were selected of $\pm$1V amplitude and 10ms width.
\label{MNN} }
 \end{center}
\end{figure}

In Fig.~\ref{MNN}(a) we show a simple MNN consisting of two memristive systems mimicking the behavior of synapses and three electronic neurons (two input and one output) whose role is to monitor their inputs and send (in both
forward and backward directions) signals of a
given intensity and shape~\cite{pershin10c}. This network belongs to a wide class of asynchronous (stochastic) neural networks \cite{Kondo92a}, namely, networks with probabilistic neuron firing. In Ref.~\cite{pershin10c} we have implemented the behavior of the electronic neurons
using an analog-to-digital converter and a microcontroller, so that once the input voltage exceeds a threshold voltage, both forward (of positive polarity) and backward (of negative polarity)
pulses are generated whose amplitude is constant, but pulse separation varies according to
the amplitude of the input signal. The basic synaptic activity of two electronic synapses ($S_1$ and $S_2$ in Fig.~\ref{MNN}(a)) has been achieved using memristor emulators~\cite{pershin09b,pershin10c,pershin10d} pre-programmed with equations of a threshold-type first-order memristive system \cite{pershin09b}:
\begin{eqnarray}
I&=&x^{-1}V_M, \label{eq3} \\ \dot x&=&\left(\beta V_M+0.5\left(
\alpha-\beta\right)\left[ |V_M+V_t|-|V_M-V_t| \right]\right)
\nonumber\\ & &\times \theta\left( x/R_1-1\right) \theta\left(
1-x/R_2\right) \label{eq4},
\end{eqnarray}

where $I$ and $V_M$ are the current through and the voltage drop on the device, respectively, and
$x$ is the internal state variable playing the role of memristance, $R=x$,
$\theta(\cdot)$ is the step function, $\alpha$  and $\beta$
characterize the rate of memristance change at $|V_M|\leq V_t$ and
$|V_M|> V_t$, respectively,
$V_t$ is a threshold voltage, and  $R_1$ and $R_2$ are limiting
values of the memristance $R$. In Eq. (\ref{eq4}), the $\theta$-functions
symbolically show that the memristance can change only between
$R_1$ and $R_2$.~\footnote{In the actual microcontroller's software
implementation, the value of $x$  is monitored at each time step
and in the situations when $x<R_1$ or $x>R_2$, it is set equal to
$R_1$ or $R_2$, respectively. In this way, we avoid situations
when $x$ may overshoot the limiting values by some amount and thus
not change any longer because of the step function in Eq.
(\ref{eq4}).}

When $\alpha=0$ and $V_t$ is above the amplitude of the
forward (pre-synaptic) pulse but does not exceed its doubled amplitude, the change in memristance ($\Delta R$) is possible only when forward (from $N_1$ or $N_2$) and
backward (from $N_3$) pulses overlap. This can be clearly seen if we rewrite Eq. (\ref{eq4}) at $\alpha=0$ (omitting $\theta$-functions) as
\begin{equation}
\dot{x}= \left\{ \begin{array}{ll}  0 & \textnormal{if} \;\;\; |V_M|\leq V_t, \\
\textnormal{sgn}(V_M)\beta (|V_M|-V_t) &  \textnormal{otherwise.} \end{array} \right. \label{eq4a}
\end{equation}
Fig. \ref{MNN}(b) shows the dependence of $\Delta R$ on the relative pulse timing.
The maximum change of memristance occurs when the pulse overlap is perfect ($\delta t=0$).

\begin{figure}[t]
 \begin{center}
\includegraphics[angle=0,width=6.5cm]{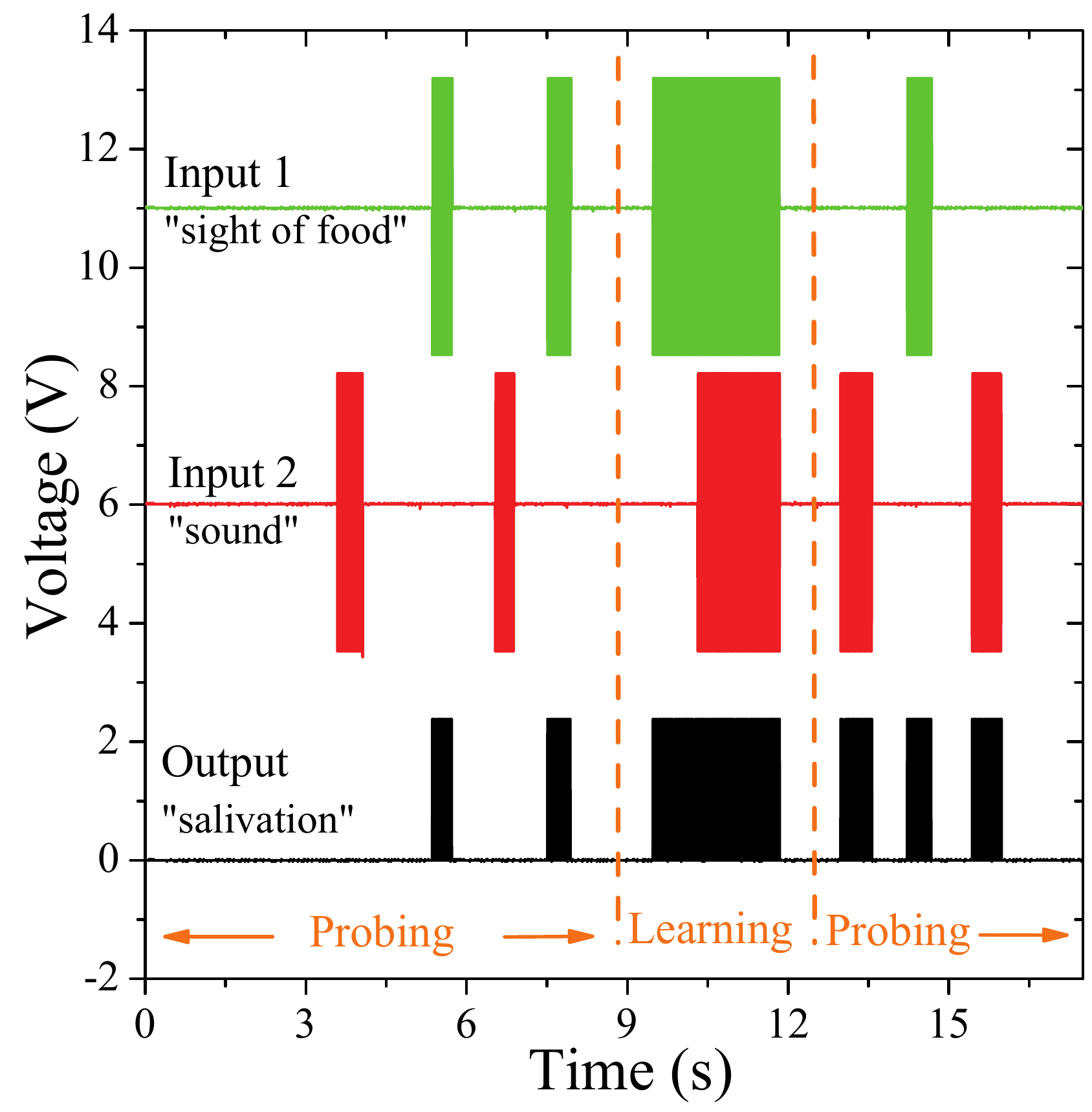}
\caption{(Color online) Experimental demonstration of the associative memory with
memristive neural networks. In this experiment, a simple neural network shown in Fig. \ref{MNN}(a)
was realized. The first "probing" phase demonstrates that, initially, only a signal from N$_1$ neuron activates
the output neuron. The association of the Input 2 signal with the Output develops in the "learning phase" when
N$_1$ and N$_2$ neurons are simultaneously activated. In this case, a signal at the Input 1
excites the third neuron that sends back-propagating pulses of a negative
amplitude. These pulses, when applied simultaneously with forward propagating pulses from the Input 2 to the second memristive synapse S$_2$ cause it to learn. The final "probing" phase demonstrates that signals from both N$_1$ and N$_2$ activate the output neuron.
A detalied description of the experiment is given in Ref.  \cite{pershin10c}.
Reprinted from Ref. \cite{pershin10c} with modifications, copyright (2010), with permission from Elsevier. \label{MNN1} }
 \end{center}
\end{figure}

This memristive synapse allows the implementation of the simplest learning rule: Hebbian learning. In fact, if we allow both input electronic neurons to fire at random times, potentiation of the conditioned memristive system
$S_2$ would occur only when the signal of the output neuron $N_3$ overlaps in time with the signal coming from neuron $N_2$.
 In this case, the second synaptic connection $S_2$ is driven from a high-resistance
to a low-resistance state since the voltage drop across $S_2$ exceeds the
threshold value due to the simultaneous excitation of both input neurons (the first one
exciting in turn the output neuron).
This is shown in Fig.~\ref{MNN1} where a detail of this process is described.

\subsection{Spike-timing-dependent plasticity}
However, biological synapses show a much more complicated plasticity (with time-resolved properties) \cite{Levy83a,Markram97a,Bi98a,Froemke02a} than the above simple Hebbian rule seems to suggest. In fact, when a post-synaptic signal reaches the synapse {\it before} the action potential of the pre-synaptic neuron, the synapse shows long-term depression (LTD), namely its strength decreases (smaller connection between the neurons) depending on the time difference between the post-synaptic and the pre-synaptic signals. Conversely,
when the post-synaptic action potential reaches the synapse {\it after} the pre-synaptic action potential, the synapse undergoes a long-time potentiation (LTP), namely the signal transmission between the two neurons increases in proportion to the time difference between the pre-synaptic and the post-synaptic signals. These general features can be implemented using different types of memristive systems.

\subsubsection{STDP with first-order memristive systems}

\begin{figure}[tb]
\begin{center}
\includegraphics[angle=0,width=6.5cm]{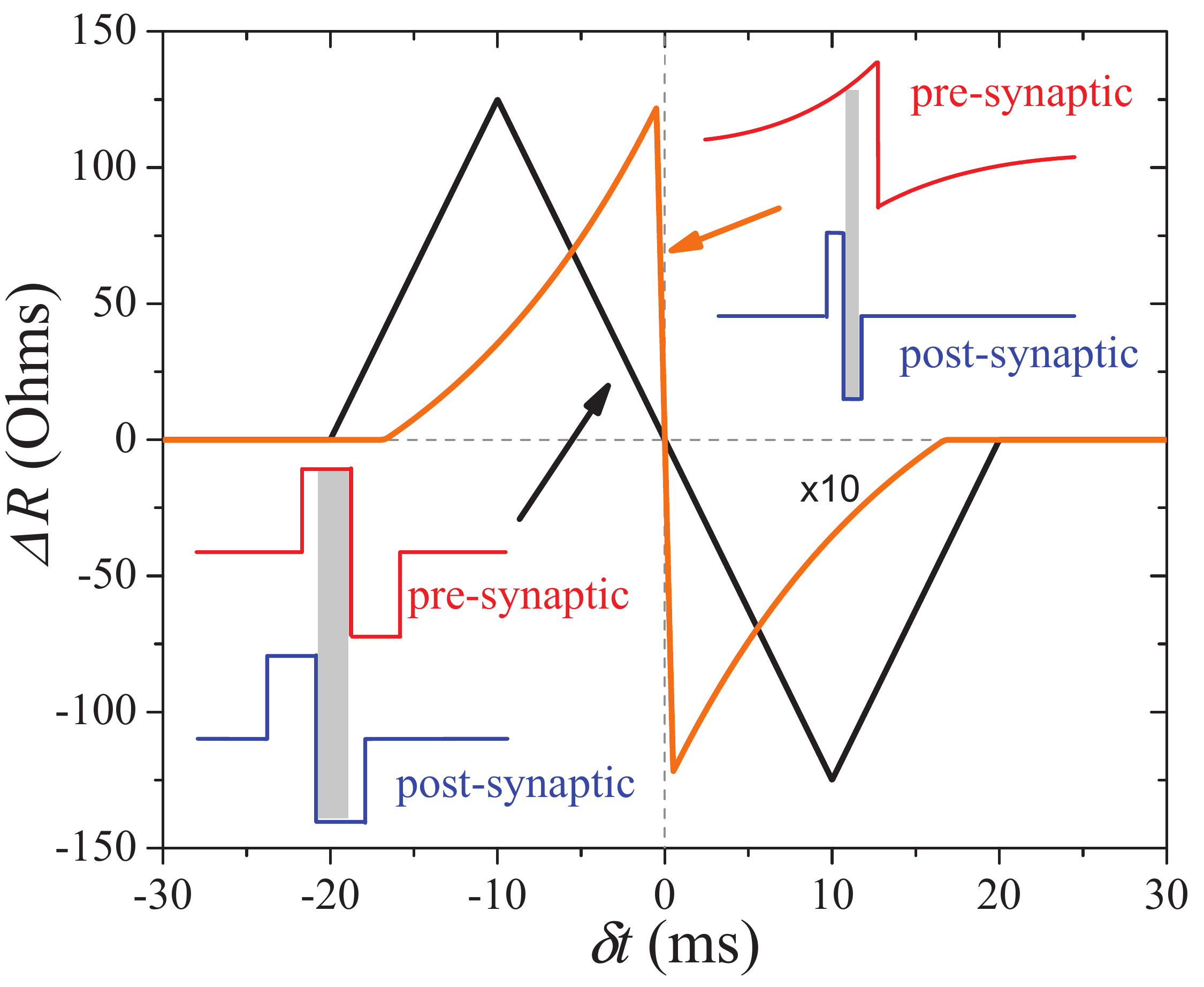}
 \caption{(Color online) Change in memristance $\Delta R$ of a first-order memristive system induced by pair of double pulses (black line) and combination of a non-linear pulse with a double pulse (orange line). The pulse form is shown schematically
 in the insets (not to scale). The grey rectangles designates the time window in which the memristance change occurs.
 These curves were obtained as a numerical
 solution of Eqs. (\ref{eq3}, \ref{eq4}) at $\alpha=0$, $\beta=25$kOhms/(V$\cdot$s), $V_t=1.5$V, $R(t=0)=2$kOhms, $R_1=1$kOhms and $R_2=10$kOhms.
 For black curve, each double pulse is composed by a 10ms-wide 1V amplitude pulse immediately followed by 10ms-wide -1V amplitude pulse.
 For orange curve, the rectangular double pulse is composed by a 0.5ms-wide 1V amplitude pulse immediately followed by 0.5ms-wide -1V amplitude pulse. The non-linear pulse is described by $V_p=V_0 \exp (t/t_0) \theta(-t)-V_0 \exp (-t/t_0) \theta(t)$ with $V_0=1.5$V and $t_0$=15ms.
 \label{stdp_pulse12} }
 \end{center}
\end{figure}

The spike-timing-dependent plasticity with first-order memristive systems can be achieved by application of bipolar pulses corresponding to pre-synaptic and post-synaptic action potential pulses. In order to achieve STDP on time scales similar to those encountered in biological synapses, the pulse width should be of the order of 20ms since the memristance modification with this type of memristive systems is possible only during pulses overlap.
Fig. \ref{stdp_pulse12} demonstrates modification of memristance induced by a pair of rectangular double pulses and also shows the effect of a non-linear and rectangular double pulse on the memristance $R$. In both cases we observe memristance changes only in a finite time window. Although rectangular double pulses are easier to generate, the use of a non-linear pulse demonstrates a closer similarity of memristance change to the variation of synaptic strength in biological systems.

We also note that the approach to synaptic plasticity using wide pulses may require reconsideration of rules governing the neural network operation. For example, it is not quite clear how to model multiple pre-synaptic pulses that are fired with a time interval shorter than their pulse length. In this publication, we are not going to focus on such issues and discuss only possible realizations of STDP with memristive synapses.

\subsubsection{STDP with second-order memristive systems}

Having discussed spike-timing-dependent plasticity with first-order memristive systems, let us consider
STDP with higher-order memristive systems. The advantage of using higher-order memristive systems is
related to their multiple state variables that can  make the operation of an artificial synapse closer to the operation of its
biological counterpart. In particular, we will focus on second-order memristive systems whose internal
state, by definition \cite{chua76a,diventra09a}, is described by two state variables.
In our model given below, the second state variable $y$ is used to track the time separation
between pre-synaptic and post-synaptic action potential pulses.

We consider a second-order memristive system described by the following equations
\begin{eqnarray}
I&=&x^{-1}V_M, \label{stdp_memr0} \\ \dot x&=&\gamma \left[ \theta (\tilde{V}_M-1) \theta (\tilde{y}-1) + \right. \nonumber \\
& & \left. \theta (-\tilde{V}_M-1) \theta (-\tilde{y}-1) \right]y,    \label{stdp_memr1} \\
\dot y&=&\frac{1}{\tau} \left[ -V_M \theta (\tilde{V}_M-1) \theta (1-\tilde{y}) - \right. \nonumber \\
& & \left. V_M\theta (-\tilde{V}_M-1) \theta (\tilde{y}+1) -y \right],
 \label{stdp_memr2}
\end{eqnarray}
where $x$ and $y$ are internal state variables, $\gamma$ is a constant, $\tilde{V}_M=V_M/V_t$, $\tilde{y}=y/y_t$ $V_t$ is a threshold voltage, $y_t$ is a threshold value of $y$, $\tau$ is a constant defining the time window of STDP. It is assumed that short (e.g., $\sim 1$ms width) pre-synaptic and post-synaptic square pulses of the same polarity are applied to the second-order memristive system. According to Eq. (\ref{stdp_memr1}), the memristance can change when $|y|\geq y_t$. The change of $y$ is described by Eq. (\ref{stdp_memr2}) whose right-hand side contains excitation terms involving $\theta$-functions and a relaxation term $-y/\tau$. Therefore, after being excited, the decay of the variable $y$ occurs with a decay constant $\tau$. The particular combination of $\theta$-functions in this equation defines the excitation rules: i) the excitation is possible only when $|V_M|>V_t$ and ii) the variable $y$ excited by a certain polarity of the voltage applied to the memristive system ($V_M$ is given by a difference of pre-synaptic and post-synaptic potentials) can not be re-excited by a pulse of opposite polarity if $|y|> y_t$. We also note that the change in memristance described by Eqs. (\ref{stdp_memr0}-\ref{stdp_memr2}) is unconstrained as we are interested only in small changes in $R$. The constraints on the minimal and maximum values of $R$ can be introduced as it is done in Eqs. (\ref{eq3}-\ref{eq4}).

\begin{figure}[tbh]
 \begin{center}
 \includegraphics[angle=0,width=6.5cm]{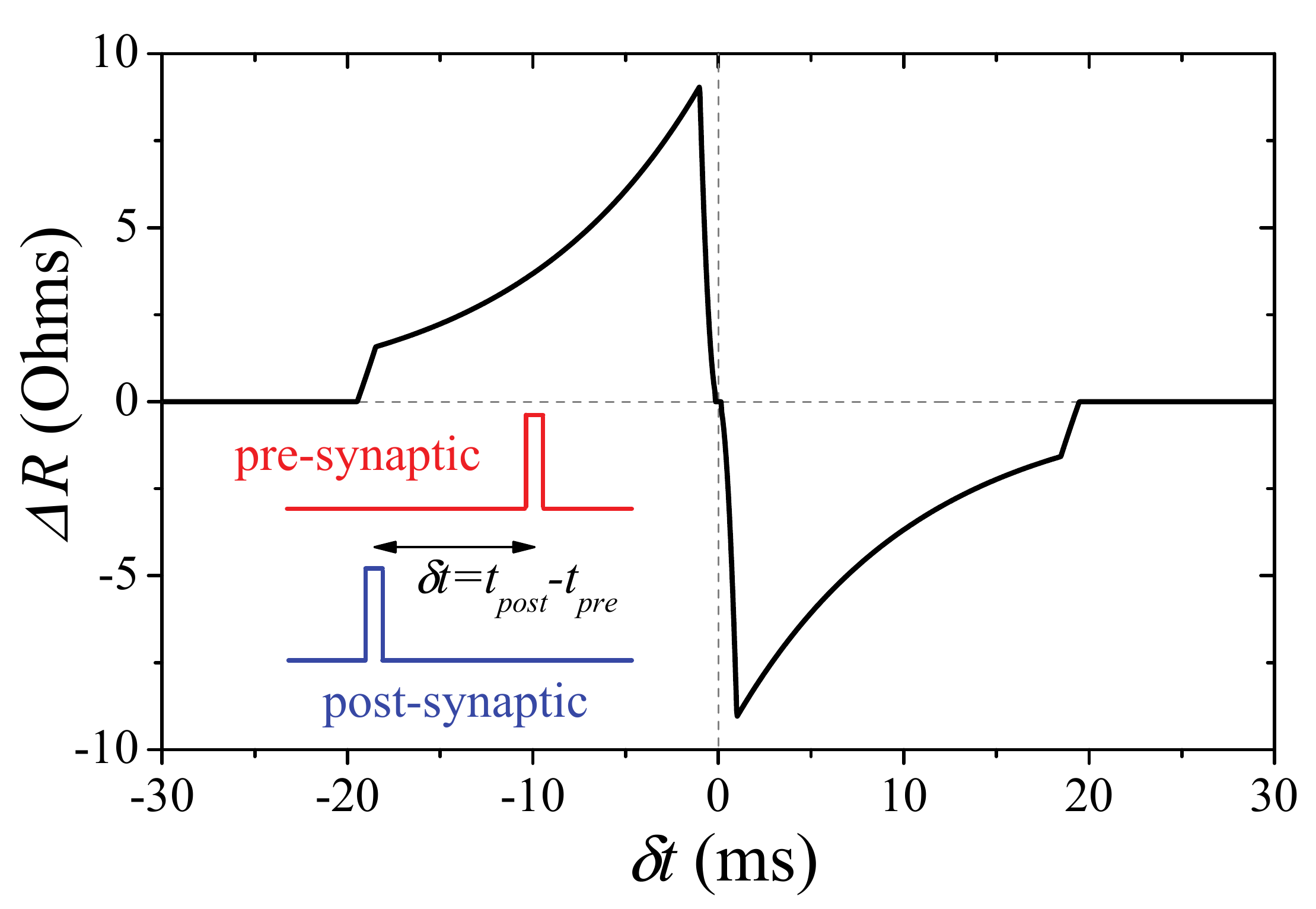}
 \caption{(Color online) Change in memristance $\Delta R$ of the second-order memristie system described by Eqs.~(\ref{stdp_memr0}-\ref{stdp_memr2}) induced by a pair of rectangular pulses separated by a time interval $\delta t$. The pulse form is schematically shown in the inset. This plot was obtained as a numerical solution of Eqs. (\ref{stdp_memr0}-\ref{stdp_memr2}) with the following set of parameters: $\gamma=10^5$C$^{-1}$, $\tau=1$ms, $V_t$=0.9V, $y_t=15$mV, $y(t=0)=0$, $x(t=0)=2$kOhms. The applied square pulses were of 1V amplitude and 1ms width.
 \label{stdp_pulse3} }
 \end{center}
\end{figure}

A numerically calculated change in memristance as a function of the time interval between the pulses is shown in Fig. \ref{stdp_pulse3}. This plot is very similar to synaptic weight changes observed in experiments with biological synapses \cite{Levy83a,Markram97a,Bi98a,Froemke02a} thus lending support to this type of model.

The use of second-order memristive systems in memristive neural networks is very promising since neuron's firing can be implemented simply  by short single rectangular pulses with no additional hardware, which is instead needed, for example, to obtain this type of functionality with only first-order memristive systems (as in Ref. \cite{jo10a}). However, such solid-state second-order memristive systems need to be developed, even though their implementation in memristor emulators~\cite{pershin09b,pershin10c,pershin10d} is straightforward. Moreover, we would like to mention that Eqs. (\ref{stdp_memr0}-\ref{stdp_memr2}) provide one of the simplest possible models of second-order memristive systems exhibiting STDP. For instance, abrupt $\theta$-functions entering these equations can be replaced by sigmoidal functions thus providing a smoother analytical behavior.

\section{Logic gates and arithmetics with memory circuit elements}

\begin{figure}[tb]
 \begin{center}
\includegraphics[angle=0,width=6.0cm]{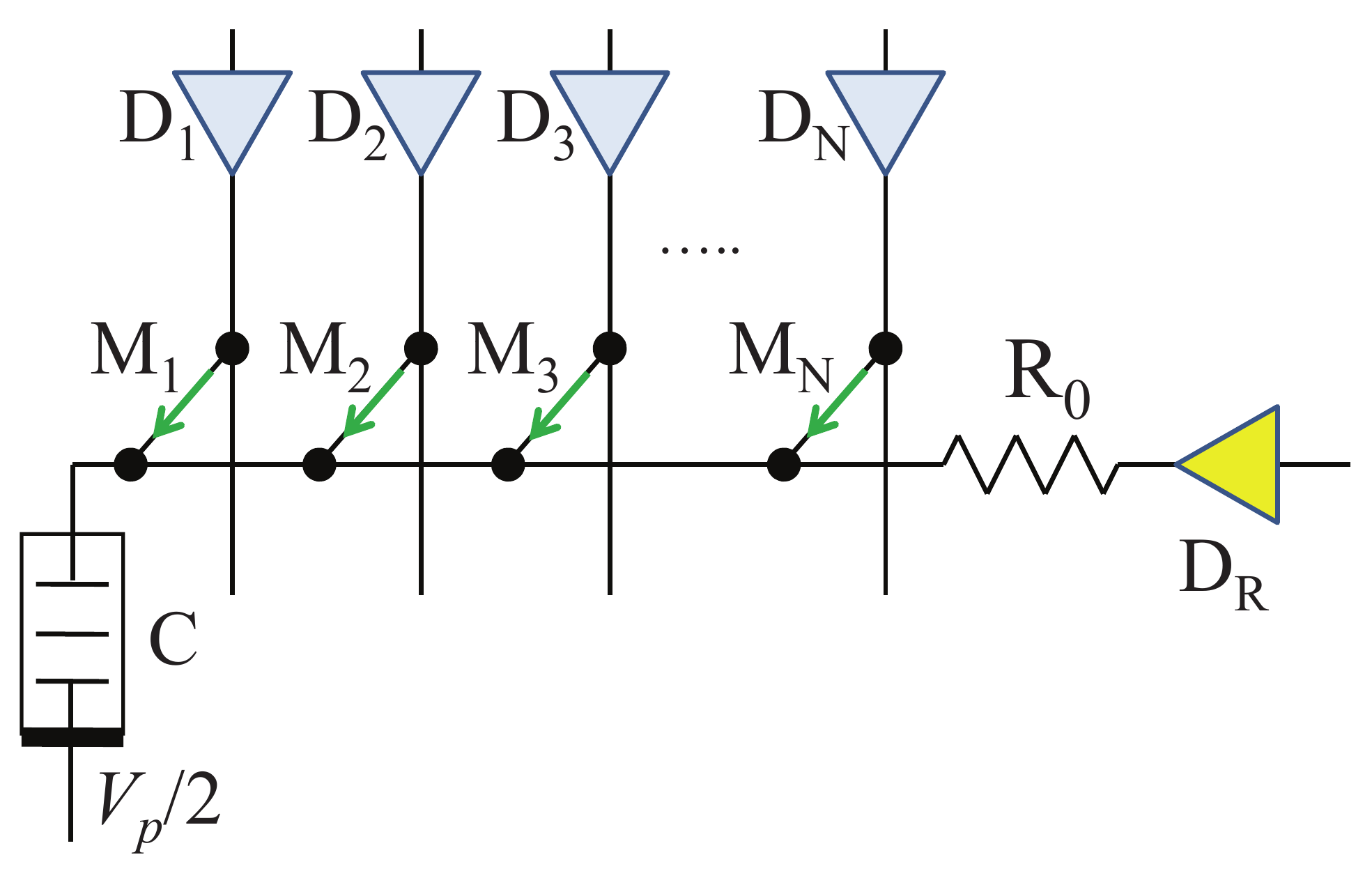}
\caption{(Color online) Electronic circuit for Boolean logic and
arithmetic operations. In this circuit, an array of $N$ memristive systems
M$_i$, memcapacitive system C and resistor R$_0$ are connected to a common
(horizontal) line. The circuit operation involves charging the
memcapacitive system $C$  through input memristive systems and discharging it
through the output ones. Because of the circuit symmetry, each
memristive system can be used as input or output. From the opposite (top)
side, memristive systems are driven by 3-state (0V, $V_p$, not connected)
drivers. The right 4-state (0V, $V_p/2$, $V_p$, not connected)
driver D$_\textnormal{R}$ connected through a resistor to the
common line is used to initialize memristive systems and memcapacitive system.
This scheme employs threshold-type bipolar memristive systems assuming
that the application of positive voltage to the top memristor electrode and
0V to the bottom electrode switches memristive system from low resistance
state (1 or ON) to the high resistance state (0 or OFF). It is also
assumed that the memristor systems' threshold voltage $V_{t}$ is between
$V_p/2$ and $V_p$. \label{ar_scheme} }
 \end{center}
\end{figure}

In this section, we discuss logic and arithmetic
operations performed by the circuit shown in
Fig. \ref{ar_scheme} which comprises an array of memristive systems, a memcapacitive system, a load resistor and drivers. In this type of application, each memristive system is used in the
``digital" mode of operation, namely only one bit of information is
encoded in the memristor's state. We call ``1" (ON) the state of lower resistance and ``0" (OFF) that of higher resistance. The operation of the circuit sketched in Fig. \ref{ar_scheme}
relies on charging a memcapacitive system through input memristive system and
subsequent discharging through the output ones. In this way, the
basic logic operations (NOT, OR, AND) as well as many other
operations (XOR, multi-input OR, AND, etc.) are
realized in a much simpler fashion than in existing approaches based on memristive systems only
\cite{Lehtonen09a,borghetti10a}. More complex
operations (e.g., addition) can be performed by sequences
of the elementary logic operations mentioned above.

Using four
memristor emulators \cite{pershin10d,pershin10c,pershin10b}, we
have built the circuit described in Fig. \ref{ar_scheme} and used it to
demonstrate the full set of basic Boolean logic operations as well
as addition of two one-bit numbers. The extension of such a
circuit to operations with multi-bit numbers is straightforward and is not reported here. In fact, the addition of two
$n$-bit numbers in this architecture requires $3n+1$
memristive systems ($2n$ memristive systems are used to store initial values, $n$
memristive systems are reserved for the calculation result and 1 memristive system
is used as a carry flag). In the circuit shown in
Fig. \ref{ar_scheme} the initial values are not destroyed and can
be reused later. The circuit uses three-state
drivers (with voltage states: 0V, $V_p$, not-connected) connected to memristive systems, and
four-state driver (0V, $V_p/2$, $V_p$, not-connected) connected to
the resistor R$_0$.
\begin{figure}[t]
\begin{center}
\includegraphics[angle=0,width=6.5cm]{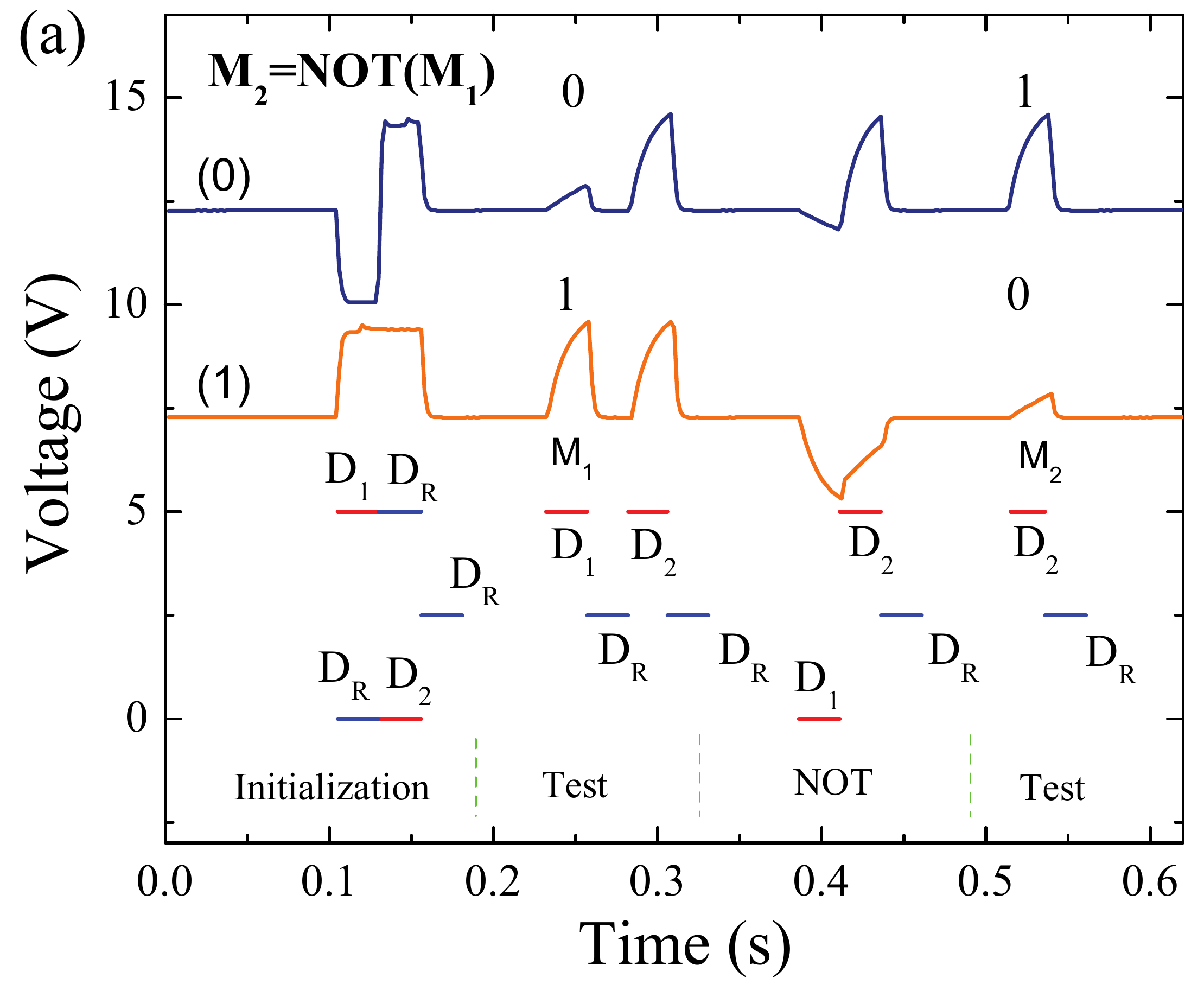}
\includegraphics[angle=0,width=6.5cm]{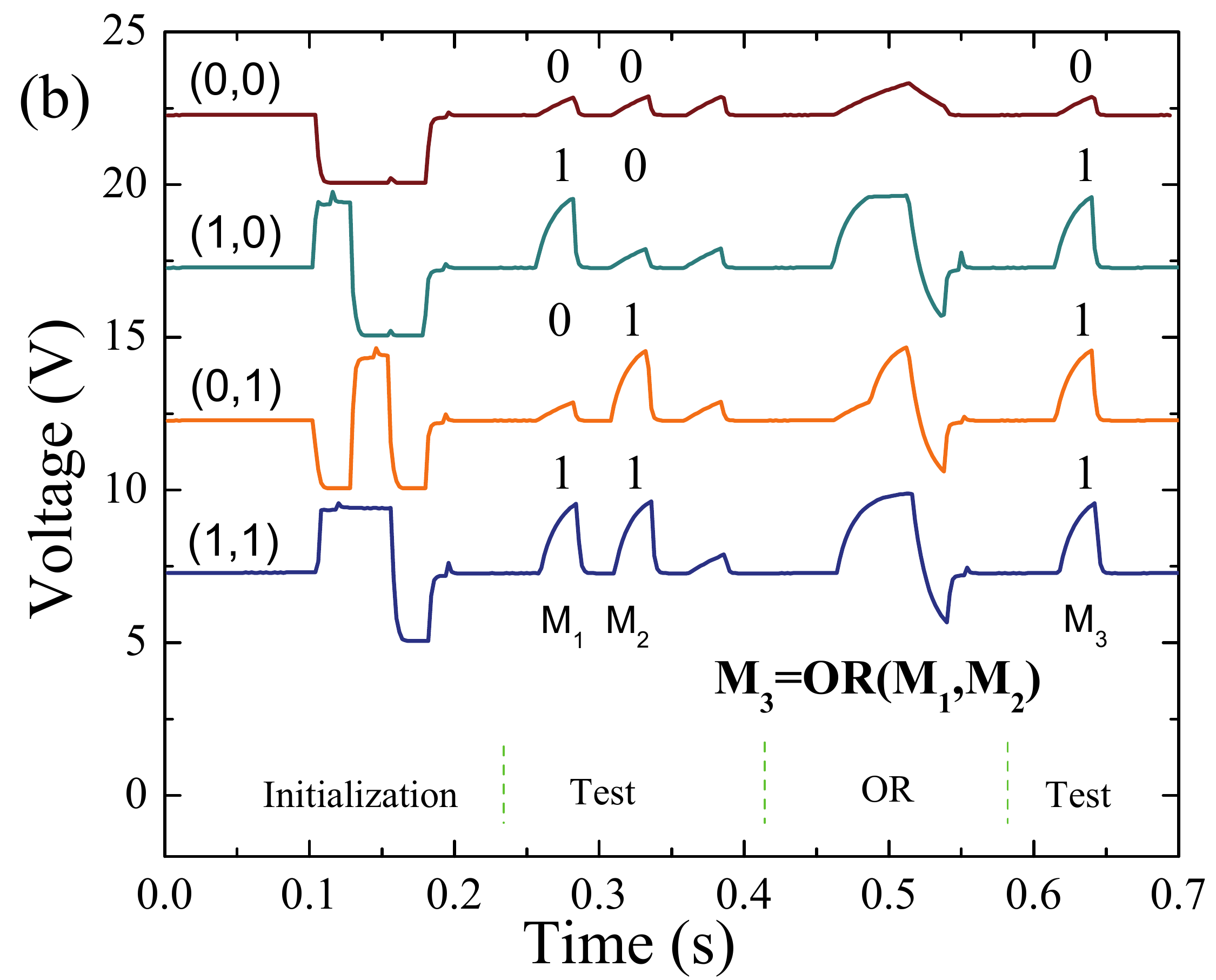}
\includegraphics[angle=0,width=6.5cm]{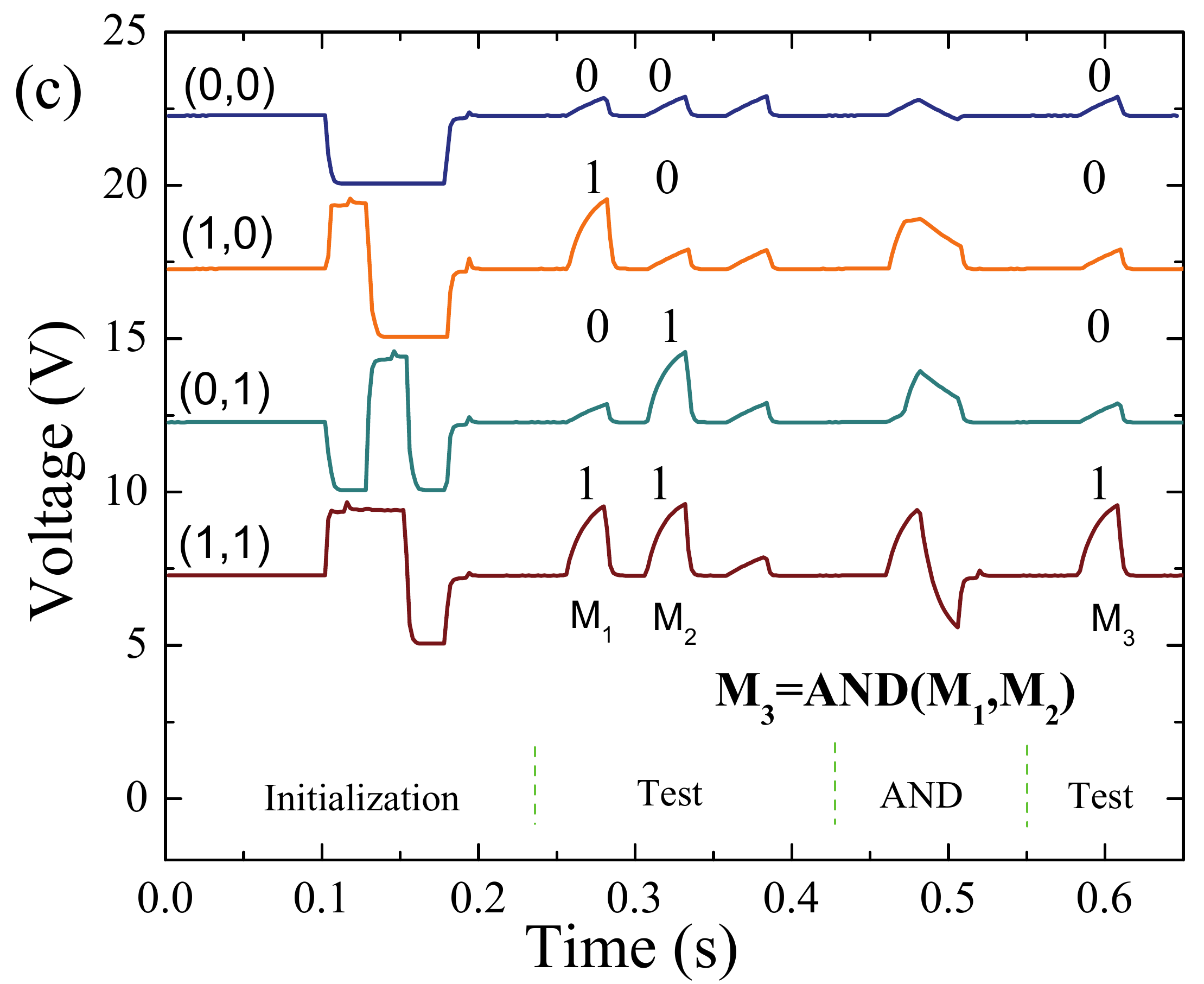}
\caption{(Color online)
Experimental realization of the basic
logic gates with memristive systems based on the circuit shown in Fig.
\ref{ar_scheme} with a capacitor $C=10\mu$F instead of a
memcapacitive system. In our experimental circuit, the role of
memristive systems is played by memristor emulators
\cite{pershin10d,pershin10c,pershin10b} governed by the
threshold-type model of Eqs.~(\ref{eq3}) and~(\ref{eq4}) with the following
set of parameters: $V_{t}=4$V, $\alpha=0$,
$\beta=62$MOhms/(V$\cdot$s). Each plot shows
several measurements of the voltage between the top plate of the capacitor (the common line defined in
the caption of Fig. \ref{ar_scheme}) and ground taken for different possible
states of input memristive systems. The curves are labeled according to the input values. The detailed information about
applied pulse sequences is given in the text. In the bottom of (a) we also show an example of voltages applied to the drivers to generate the top voltage plot (0). Here, we show only voltages applied by drivers, namely, the voltages when a driver is not in "not connected" state.
The voltages on the drivers in (a) are in absolute values while the other curves were vertically displaced
for clarity.
 \label{basic_gates} }
 \end{center}
\end{figure}

Although memcapacitor (as well as meminductor) emulators can also be built using simple electronic schemes~\cite{pershin10b,pershin11a}, their
models are not yet well developed. Moreover, the noise reduction is a very important task in experiments with such schemes~\cite{pershin10b}. Therefore, in our circuit realization we have employed
a usual $10\mu$F capacitor in place of the memcapacitive system. We will later discuss the advantages of replacing the regular capacitor with a memcapacitor.

\begin{figure}[h]
 \begin{center}
\includegraphics[angle=0,width=6.5cm]{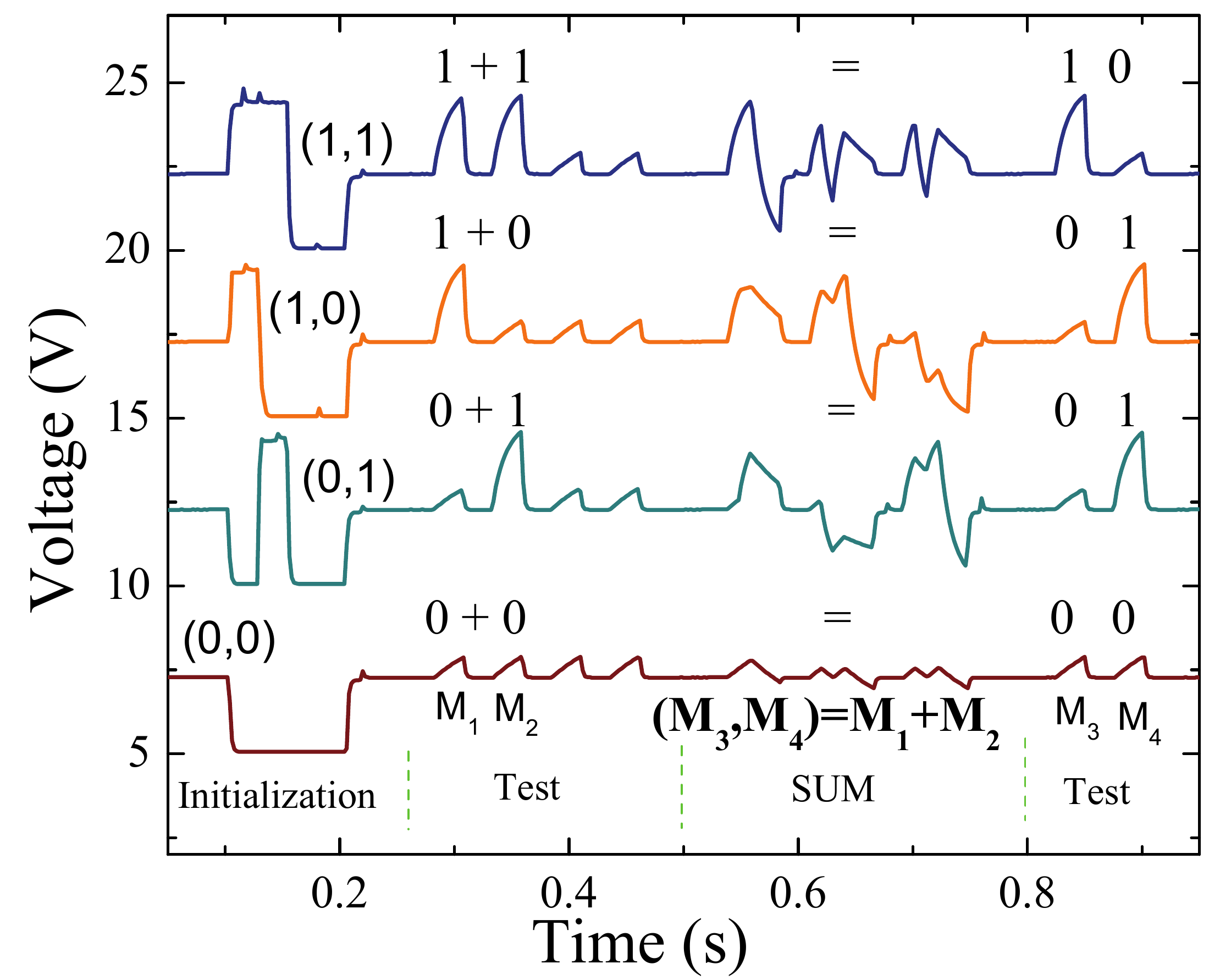}
\caption{(Color online) Summation of two one-bit numbers performed
as two logic operations: M$_3$=AND(M$_1$,M$_2$) and
M$_4$=XOR(M$_1$,M$_2$). The voltage curves corresponding to different
values of input numbers have been vertically displaced for clarity.
\label{sum} }
 \end{center}
\end{figure}

In order to understand the circuit operation, we first note that
the circuit employs threshold-type bipolar memristive systems such as the TiO$_2$
memristive systems recently demonstrated  experimentally \cite{Yang08a}. To represent this situation we have implemented the equations~(\ref{eq3}) and~(\ref{eq4}) in our
memristor emulator~\cite{pershin09b,pershin10c,pershin10d}. In this case, the change of
memristance occurs only if the applied voltage
exceeds the threshold voltage $V_t$ of the device. In addition, for successful
circuit operation, the value of $V_t$ should be selected between
$V_p/2$ and $V_p$. In this case, the charging of $C$ from the
completely discharged state will never alter the value of input
memristive systems (since during this process voltages applied to input
memristive systems do not exceed $V_p/2$).

The circuit operation consists of a
sequence of the following elementary operations: 1) capacitor
initialization (by application of $V_p/2$ voltage pulse by
D$_\textnormal{R}$ driver), 2) initialization of memristive systems' states (by
application of 0 or $V_p$ voltage pulse by D$_\textnormal{R}$
driver simultaneously with $V_p$ or 0 voltage pulse by one of
D$_1$-D$_4$ drivers), 3) capacitor charging through an input memristive system
(by application of 0 or $V_p$ voltage pulse by one of the D$_1$-D$_4$
drivers), and 4) capacitor discharge through an output memristive system (by
application of $V_p$ or 0 voltage pulse by one of D$_1$-D$_4$
drivers). Note, that in the operations used by us the current can flow only through one memristive system at any one time. Namely, after the capacitor has been charged through the input
memristive system, the latter is disconnected (its driver is switched into "not connected" state) and the output memristive system is connected to allow capacitance discharge.~\footnote{Note that the switching of driver states can be done very fast.} The parameter values specific to our circuit realization are:
 $V_p=5$V, $V_{t}=4$V, and 25ms is the width of initialization and reset pulses. An example of sequences of voltage drivers necessary to create the
 NOT gate is shown in Fig. \ref{basic_gates}(a).
Having discussed elementary circuit operations, let us now
turn to implementation of basic Boolean operations.

{\it NOT gate} - M$_2$ = NOT(M$_1$) - 4 steps - This operation is realized according to the following sequence of operations (see
also Fig. \ref{basic_gates}(a)): set M$_2=1$, reset C, charge C
through M$_1$ by 0V pulse of a finite width applied by D$_1$,
discharge C through M$_2$ by $V_p$ pulse applied to D$_2$. The
main idea is that the capacitor C can only be charged to a
sufficiently high voltage to switch M$_2$ to zero if M$_1=1$.
Therefore, if M$_1=1$ then M$_2\rightarrow 0$ and if M$_1=0$ then
M$_2=1$. This is the logical NOT operation.

Going into measurement
details, in the initialization phase (Fig. \ref{basic_gates}(a))
we set M$_2=1$ and  M$_1=0$ (top curve (0)) or M$_1=1$ (curve (1))
by application of 25ms pulses by D$_1$, D$_2$ and D$_\textnormal{R}$ drivers as described above. In the first testing phase, we monitor the memristance of M$_1$ and M$_2$ by consecutive application of positive 25ms pulses by
D$_1$ and D$_2$ drivers to corresponding memristive systems as well as
capacitor reset pulses. In this way, $C$ is charged through the
memristive systems and the charging rate provides information about
memristive systems' states. Then, we reset C, charge C through M$_1$ by
25ms 0V pulse, discharge C through M$_2$ by 25ms $V_p$ pulse. In
the final testing phase, the value of $M_2$ is monitored via the
rate of the capacitor charging, again via application of a 25ms $V_p$ pulse to M$_2$. If the voltage across the capacitor does not change
much then the state of M$_2$ is $0$: high-resistance state and thus low current flows to the capacitor plates. Conversely, if the
voltage across the capacitor does change faster to $V_p$  then the state of M$_2$ is $1$: low-resistance state and thus more charges accumulate on the capacitor plates. For the purpose of clarity,
the sequence of applied pulses is shown in the bottom of Fig.
\ref{basic_gates}(a) in absolute scale for the realization of the top sequence of Fig.
\ref{basic_gates}(a). All other curves are shifted for clarity.

{\it OR gate} - M$_3$ = OR(M$_1$,M$_2$) - 5 steps - This is implemented by the
following sequence: M$_3=0$, reset C, charge C through M$_1$ by
$V_p$ pulse, charge C through M$_2$ by $V_p$ pulse, discharge C
through M$_3$ by 0V pulse. The width of all pulses in this
sequence is 25ms. Figure \ref{basic_gates}(b) shows four voltage lines
corresponding to different initial states of M$_1$ and M$_2$
memristive systems. The correct truth table of the OR gate is clearly
identified during the second testing phase in Fig.
\ref{basic_gates}(b).

{\it AND gate} - M$_3$ = AND(M$_1$,M$_2$) - 5 steps - The sequence for this logic gate is as follows: M$_3=0$,
reset C, charge C through M$_1$ by a short $V_p$ pulse, charge C
through M$_2$ by a short $V_p$ pulse, discharge C through M$_3$ by
0V pulse. In the AND operation, the width of short pulses is equal
to 10ms and the width of all other pulses is 25ms. The width of
short charging pulses is selected in such a way that both M$_1$
and M$_2$ memristive systems should be in the ON state in order to charge
$C$ to such a voltage so that during the next step, namely during the capacitor discharge through M$_3$, the latter is
switched into the ON state. Fig. \ref{basic_gates}(c) shows
results demonstrating correct implementation
of the AND gate.

{\it XOR gate} - M$_3$ = XOR(M$_1$,M$_2$) - 11 steps - This is realized by the
following sequence: M$_3=0$, reset C, charge C through M$_1$ by a short $V_p$ pulse, discharge C through M$_2$ by a short $0$V pulse, charge C through M$_1$ by a short $V_p$ pulse, discharge C through M$_3$ by 0-V pulse, reset C, charge C through M$_2$ by a short $V_p$ pulse, discharge C through M$_1$ by a short $0$V pulse, charge C through M$_2$ by a short $V_p$ pulse, discharge C through M$_3$ by 0-V pulse. Similarly to the AND gate, the duration of short pulses is 10 ms and that of all other pulses is 25 ms.

{\it Addition of two 1-bit numbers} - 16 steps - Next, we consider the addition of two one-bit
numbers that we encode in M$_1$ and M$_2$ memristive systems. The result
of addition is saved in M$_3$ and M$_4$ memristive systems. It is easy to
check that the addition can be performed with the sequence M$_3$=AND(M$_1$,M$_2$)
and M$_4$=XOR(M$_1$,M$_2$). In Fig. \ref{sum} we show
experimentally measured signal patterns that prove the above operation. We emphasize that an approach suggested previously \cite{Lehtonen09a}
would require 87 steps to perform the same operation.

At this point, we would like to re-emphasize that the extension of our
scheme to the addition of $n$-bit numbers is straightforward. In
particular, the $i$-th bit of the result will be given by a sum of the two
$i$-th bits of the input numbers and the carry flag, from the summation of the $i$-1 bits. In the case of the one-bit addition considered in the present work, the
carry flag is encoded in the M$_4$ memristive system. Moreover, we want to stress that the circuit
operation can be further optimized if we use a memcapacitive system as
shown in Fig. \ref{ar_scheme} (instead of the usual capacitor implemented here). In this case, the state of the memcapacitive system can evolve when
pulses are applied to those memristive systems that store the initial values.
Therefore, some computation results can be temporarily stored in
the memcapacitive system state thus decreasing the number of computation
steps required.

Finally, we would like to mention that the crossbar architectures - formed by ``vertical" and ``horizontal" sets of wires -
typically used to fabricate nanoscale memristive systems \cite{Yang09b,Jo09a,Borghetti09a} may
allow an effective parallelization of the arithmetic and logic
operations we have described here. For instance, the application of pulses to ``vertical" wires would cause
a computation within each horizontal line in parallel. Moreover,
switching the role of horizontal and vertical wires would cause a
switch of computation direction. In this way, interesting and more complex
computational schemes can be practically realized. However, a practical realization of such a scheme
may require use of highly nonlinear memristive systems or additional switches/individual access devices in order to avoid
unwanted currents that naturally appear in a resistive network.
In addition, if unwanted currents can be eliminated or reduced, this massive parallelization may offset intrinsically slower
switching times of memristive systems ($\gtrsim 10$ns) in comparison with typically shorter switching times of, e.g., CMOS-based elements performing arithmetic operations.

\section{Quantum computation with memory elements}

It seems timely to consider the potentialities of memory elements - in particular low-dissipative meminductive and memcapacitive systems - in quantum computation,
which is based on the unitary evolution of a quantum system. Although many experimental systems  have been
suggested as quantum bits (qubits) \cite{Steane97a,Loss98a,Privman98a,Pershin03a,Clarke08a}, here we will focus on superconducting (SC) qubits \cite{Clarke08a,zagoskin07a} that currently are considered among the most promising ones. Since typical SC qubit circuits involve capacitors and inductors, memcapacitive and meminductive systems fit naturally in this application.

\begin{figure}[t]
 \begin{center}
\includegraphics[angle=0,width=8cm]{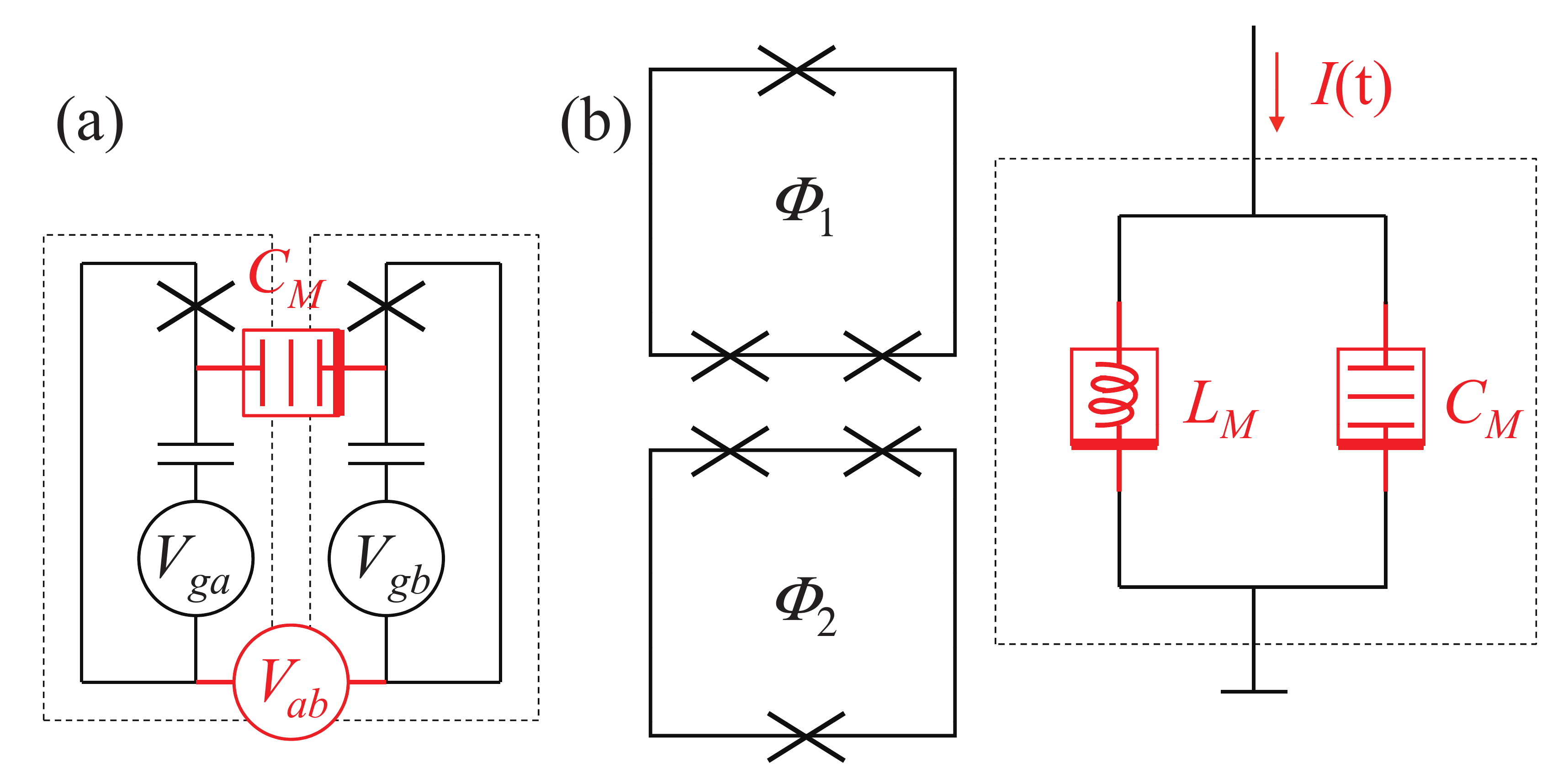}
\caption{(Color online) (a) Suggested coupling of charge qubits using a memcapacitive system $C_M$. Here,
$V_{ab}$ is a controllable voltage source used to set the state of the memcapacitive system, while
$V_{ga}$ and $V_{gb}$ are voltage sources needed to control the number of electrons across the Josephson junction represented by the cross sign. (b) Suggested coupling of flux qubits using a tank circuit with a meminductive system, $L_M$ and a memcapacitive system, $C_M$. Here,
$\Phi_1$ and $\Phi_2$ are the fluxes that are controlled by the current flowing in the tank circuit and possibly by other controlling coils. \label{qubits} }
 \end{center}
\end{figure}

{\it Field-programmable quantum computation -}
In SC quantum computation schemes, phase and charge qubits are coupled capacitively while flux qubits are coupled inductively \cite{zagoskin07a}.
Memcapacitive and meminductive systems can be used in such schemes to provide a controllable interaction between SC qubits~\footnote{The range of possible applications of memory elements in quantum computing is not limited to this situation, and we indeed anticipate many other different opportunities. For instance, a micromechanical resonator embedded into a DC SQUID~\cite{poot10a} results in an effective meminductive system. In addition, a lagging of SC qubit's due to memory effects in inductance was recently discussed \cite{shevchenko08a}.}. For example, let us consider two charge qubits \cite{Bouchiat98a,Nakamura99a} interacting via a usual capacitive coupling. The corresponding coupling term has the form \cite{zagoskin07a}
\begin{equation}
H_{int}=\frac{e^2}{2C_x}\sigma_a^x\sigma_b^x,\label{Hint}
\end{equation}
where $e$ is the elementary charge and $\sigma_{a(b)}^x$ are the $x$-component Pauli matrices acting on the wave function of the corresponding qubit in the reduced Hilbert subspace. We can now replace the capacitor $C_x$ by a memcapacitive system $C_M$ and insert a controllable voltage source $V_{ab}$ between the qubits (see Fig.~\ref{qubits}(a)). Alternatively, $V_{ab}$ can be directly connected in parallel with the memcapacitive system. This last scheme, however, may induce large currents in the two qubits if the qubits circuits are grounded thus perturbing them  incoherently. Either way, the value of the memcapacitance - that replaces $C_x$ in Eq. (\ref{Hint}) - can be {\it pre-set} and the interaction Hamiltonian will depend on this pre-set value of capacitance. For two interacting qubits the magnitude of the interaction strength is not very important as it can be compensated by the interaction time. However, if we consider $N$ simultaneously interacting qubits then a variation of coupling between two of them will result in absolutely different interaction Hamiltonians thus leading to a different system evolution. Quantum computation algorithms will thus benefit from such novel quantum hardware functionality because of the many (practically infinite) interaction schemes that can be implemented
within a single device architecture.

In the case of flux qubits the latter can be controlled, via mutual inductance, by a tank circuit consisting of a meminductive system with a regular capacitor or possibly a memcapacitive system as shown in Fig.~\ref{qubits} (b). The tank circuit may be current- or voltage-controlled and by setting the different values of meminductance, $L_M$ and memcapacitance, $C_M$, its frequency $\omega=1/\sqrt{L_MC_M}$ changes thus leading to a different interaction
Hamiltonian between the two qubits. Again, in the case of $N$ interacting qubits this leads to a large number of time evolutions and thus different computation schemes. Because of the similarity to classical field-programmable gate arrays, we could term the above two computation architectures as ``field-programmable quantum computation". Some work is however needed to realize such an architecture practically, in particular, to avoid additional decoherence effects that can be introduced by real memcapacitive/meminductive elements \cite{zine04a,chang06a,driscoll09a,Lai09a,martinez10a,martinez11a}.

\section{Conclusions}

In conclusion, we have addressed all three different paradigms of computation - neuromorphic (analog), digital and quantum - and shown the potential the three classes of memory elements, namely
memristive, memcapacitive and meminductive systems offer in all three cases. In particular, we have discussed memristive neural networks,
classical logic and arithmetic operations with memristive and memcapacitive systems, and field-programmable quantum computation with meminductive and memcapacitive systems with both charge and flux qubits. Clearly, much work needs to be done to advance these applications. Important issues include, e.g., realization of high connectivity in neural networks (each neuron in the human brain has $\sim$ 7000 synaptic connections), effective parallelization  of logic and arithmetic operations,  reduction of qubit decoherence, etc. In spite of these difficulties, with the increasing research into actual materials and devices that behave as memory elements we anticipate many more ideas could be implemented in actual systems and circuits thus offering a wider range of opportunities even in hybrid computational schemes involving any two, or even all three paradigms.

\section*{Acknowledgment}
M.D. acknowledges partial support from the National Science
Foundation (DMR-0802830). We thank Ben Criger and Frank Wilhelm for getting us interested in the possible use of memory elements in quantum computation and for useful discussions.

\bibliographystyle{IEEEtran}
\bibliography{memcapacitor1}

\end{document}